\documentclass[aps,showpacs,twocolumn,superscriptaddress]{revtex4-1}
\usepackage{graphicx}
\usepackage{amsfonts}
\usepackage{amssymb}
\usepackage{amsmath}
\usepackage{natbib}
\usepackage{dcolumn}
\usepackage{bm}
\begin{document}
\title {Nuclear symmetry energy in a modified quark meson coupling model}
\author{R.N. Mishra}
\author{H.S. Sahoo}
\affiliation{Department of Physics, Ravenshaw University, Cuttack-753 003, 
India}
\author{P.K. Panda}
\author{N. Barik}
\affiliation{Department of Physics, Utkal University, Bhubaneswar-751 004, 
India}
\author{T. Frederico}
\affiliation{Departmento de F\'isica, Instituto Techn\'ologico de Aeron\'atica,
12228-900 S\~ao Jos\'e dos Campos, SP, Brazil}
%\date{\today}
\begin{abstract}
We study nuclear symmetry energy and the thermodynamic instabilities of 
asymmetric nuclear matter in a self-consistent manner by using a modified 
quark-meson coupling model where the confining interaction
for quarks inside a nucleon is represented by a
phenomenologically averaged potential in an equally mixed scalar-vector
harmonic form. The nucleon-nucleon interaction
in nuclear matter is then realized by introducing additional
quark couplings to $\sigma$, $\omega$, and $\rho$ mesons through mean-field 
approximations. We find 
an analytic expression for the symmetry energy ${\cal E}_{sym}$ 
as a function of its slope $L$. Our result establishes a linear correlation 
between $L$ and ${\cal E}_{sym}$. We also analyze the constraint on neutron star 
radii in $(pn)$ matter with $\beta$ equilibrium.
\end{abstract}

\pacs{26.60.+c, 21.30.-x, 21.65.Qr, 95.30.Tg}
\maketitle
\section{Introduction}
One of the major focuses in the study of nuclear matter has recently been 
to understand the equation of state (EOS) of asymmetric nuclear matter and the
density dependence of the nuclear symmetry energy. 
The nuclear symmetry energy is a fundamental quantity 
which determines several important properties of very small 
entities such as the atomic nuclei as well as very large objects such as neutron stars 
\cite{prakash}. 
In fact, the behavior of nuclear symmetry energy is most uncertain among all 
properties of dense nuclear matter. Furthermore, the symmetry energy is important
for modeling nuclear matter by probing the isospin part of nuclear 
interactions. Recent \cite{tili} experimental studies of 
isospin-sensitive observables in intermediate-energy 
nuclear reactions involving radioactive beams have been quite useful in 
providing some constraints on the density dependence of nuclear 
symmetry energy at subsaturation densities. The effects of symmetry energy and 
its slope on neutron star properties is an important area of study.
Another area of relevance in 
the study of asymmetric nuclear matter is the instabilities associated with possible 
liquid-gas phase transitions at subsaturation densities. Such 
liquid-gas phase transition plays an important role in the 
description of the crust of compact star matter at densities between 
$0.03$ fm$^{-3}$ and saturation density ($~0.15$ fm$^{-3}$).
Here, we would like to address these two relevant aspects in the study of 
asymmetric nuclear matter in a phenomenological model that we have used in 
our earlier work for symmetric nuclear matter.

There has been a proliferation of phenomenological models
to describe infinite nuclear matter and also properties of 
finite nuclei. These are in fact
essential steps in the development of this area of study for which realistic 
first-principles theoretical descriptions as well as adequate experimental
or observational data are not available. Variations in different phenomenological
approaches stretching from the nonrelativistic to the relativistic are
tried incorporating some further aspects of theoretical requirements in the
model. All these models are usually set in terms of parameters that are fit 
to reproduce the properties of either finite nuclei \cite{nl3} or bulk nuclear matter.
As a result, most of the models behave more or less similarly as far as the 
equation of state
is concerned around the saturation density and at zero temperature. However, when these
models are used to describe nuclear matter at subsaturation densities to explain
the liquid-gas phase transition or at high densities to explain neutron star matter,
they yield very different results. Therefore, it has been seen as essential
to incorporate some constraints related to symmetry energy and its derivatives
by using up-to-date theoretical and experimental information. Most of the 
relativistic-mean-field (RMF) models are attempts in these directions. However, in these
models nucleons are treated as structureless point objects. Therefore, as a next step
in the requirement of incorporating the quark structure of the nucleon with
meson couplings at the basic level; quark-meson-coupling (QMC) models have been proposed
\cite{guichon} and properties of nuclear matter have been studied in great detail
in a series of works \cite{ST,recent,temp,phase}. In these models nucleons are
described as a system of nonoverlapping MIT bags which interact through effective
scalar and vector meson exchanges at the quark level. However, it has been argued
that the hadronic structure described by the MIT-bag model suffers from some
theoretical inadequacy due to the sharp bag boundary in breaking chiral 
symmetry, which is a good symmetry of strong interactions within the partially 
conserved axial current (PCAC) limit.
Therefore, more sophisticated versions such as the Cloudy Bag Model (CBM) have 
been proposed for the study of hadronic structure. So, to further include this
aspect of the physics requirement, it would be more appropriate to develop a
quark-meson coupling model where nucleon structure is described by models
like the CBM instead of by MIT bags. As an alternative approach
\cite{barik,frederico89,batista}
to the CBM, the relativistic independent quark model with a phenomenologically
averaged confining potential in equally mixed scalar-vector harmonic form in 
the Dirac frame work has been used extensively with remarkable consistency
in the baryonic as well as the mesonic sector \cite{bdd}. This model has provided a
very suitable alternative to the otherwise successful cloudy bag model in 
describing hadronic structure with its static properties and various
decay properties.

We therefore proposed in our earlier work \cite{rnm} a modified quark-meson 
coupling model (MQMC), which is based on a suitable
confining relativistic independent quark potential rather than a bag 
to address the nucleon structure in vacuum as an alternative 
approach to QMC for the study of symmetric nuclear matter. 
 This attempt was not so much as to plead superiority of MQMC over QMC at
this level. It has only incorporated an essential aspect of the physics 
requirement missing in MIT-bag model to stand as an alternative to the more 
appropriate CBM. Further investigations are
necessary to check its consistency and its predictability for any
new physical features. In the MQMC model, we  have 
studied the bulk nuclear properties such as the compressibility, the 
structure of EOS, and also discussed 
some implications of chiral symmetry in nuclear matter along with the nucleon 
and nuclear $\Sigma$ term and the sensitivity of nuclear
matter binding energy with variations in the light quark mass. The results 
obtained in such a picture for symmetric nuclear matter were quite encouraging.
In the present attempt, we study the bulk properties of asymmetric nuclear 
matter 
and also the low-density instabilities of the system in such a model. To treat 
the asymmetric nuclear matter, we incorporate in our model the contribution  of 
the isovector vector meson $\rho$ in addition to those of the isoscalar scalar
meson $(\sigma)$ and isoscalar vector meson$(\omega)$ considered earlier
for symmetric nuclear matter \cite{rnm}. Such 
studies are also useful to discuss the systems such as neutron stars with 
$N \neq Z$.

A correlation between the symmetry energy ${\cal E}_{sym}$ and its slope $L$ 
has been verified recently by Ducoin {\it{et al.}} \cite{ducoin} for a set of 
effective relativistic and nonrelativistic  nuclear models. Such a study was 
based on numerical results for ${\cal E}_{sym}$ and $L$ obtained from 
different parametrizations. 
Theoretically ${\cal E}_{sym}$ and $L$ are constrained \cite{stone,chen}. 
In a recent paper, Santos {\it{et al}.} \cite{correl} have established an 
analytic relationship between these quantities. 
In this context QMC-based models have not been studied. We have made an 
attempt to set up a relationship between these two quantities analytically.

The paper is organized as follows: In Sec. II, a brief outline
of the model describing the nucleon structure in vacuum is discussed. 
The nucleon mass is then 
realized by appropriately taking into account the center-of-mass correction, 
pionic correction, and gluonic correction. The EOS is then developed. 
In Sec. III, we discuss the nuclear symmetry energy, its slope and 
incompressibility, and observe its density dependence. The thermodynamic 
instabilities of the system are analyzed in Sec. IV.
We establish the analytic relationship between ${\cal E}_{sym}$ and $L$
and discuss the results in Sec. V.

\section{Modified quark meson coupling model}

Recently, the modified quark-meson coupling model was adopted
for symmetric nuclear matter where the $NN$ interaction was realized in a 
mean-field approach through the exchange of effective $(\sigma,\omega)$ 
mesonic fields coupling to the quarks inside the nucleon \cite{rnm}. We now 
extend this model to asymmetric nuclear matter and include the contribution of 
the isovector vector meson, $\rho$, in addition to $\sigma$ and $\omega$ mesons.
In view of this, we briefly present the outlines of our approach
\cite{rnm} in the present context.

We first consider nucleons as a composite of constituent quarks confined in 
a phenomenological flavor-independent confining potential, $U(r)$ in an equally 
mixed scalar and vector harmonic form inside the nucleon \cite{rnm}, where 
\[
U(r)=\frac{1}{2}(1+\gamma^0)V(r),
\]
with 
\begin{equation}
V(r)=(ar^2+V_0),~~~~~ ~~~ a>0. 
\label{eq:1}
\end{equation}
Here $(a,~ V_0)$ are the potential parameters. The confining interaction here 
provides the zeroth-order quark dynamics of the hadron.  In the medium, the 
quark field $\psi_q({\mathbf r})$ satisfies 
the Dirac equation
\begin{equation}
[\gamma^0~(\epsilon_q-V_\omega-
\frac{1}{2} \tau_{3q}V_\rho)-{\vec \gamma}.{\vec p}
-(m_q-V_\sigma)-U(r)]\psi_q(\vec r)=0
\end{equation}
where $V_\sigma=g_\sigma^q\sigma_0$, $V_\omega=g_\omega^q\omega_0$ and
$V_\rho=g_\rho^q b_{03}$; where $\sigma_0$, $\omega_0$, and $b_{03}$ are the
classical meson fields, and 
$g_\sigma^q$, $g_\omega^q$, and $g_\rho^q$ are the quark couplings to  
the $\sigma$, $\omega$, and $\rho$ mesons, respectively. $m_q$ is the quark
mass and $\tau_{3q}$ is the third component of the Pauli matrices. In the
present paper, we consider nonstrange $q=u$ and $d$ quarks only.
We can now define
\begin{equation}
\epsilon^{\prime}_q= (\epsilon_q^*-V_0/2)~~~ 
\mbox{and}~~~ m^{\prime}_q=(m_q^*+V_0/2),
\label{eprim}
\end{equation}
where the effective quark energy, 
$\epsilon_q^*=\epsilon_q-V_\omega-\frac{1}{2}\tau_{3q} V_\rho$ and 
effective quark mass, $m_q^*=m_q-V_\sigma$. We now introduce $\lambda_q$ 
and $r_{0q}$ as
\begin{equation}
(\epsilon^{\prime}_q+m^{\prime}_q)=\lambda_q~~
~~\mbox{and}~~~~r_{0q}=(a\lambda_q)^{-\frac{1}{4}}.
\label{eq:8}
\end{equation}

The ground-state quark energy can be obtained from the eigenvalue condition
\begin{equation}
(\epsilon^{\prime}_q-m^{\prime}_q)\sqrt \frac{\lambda_q}{a}=3.
\label{eq:11}
\end{equation}
The solution of equation \eqref{eq:11} for the quark  energy
$\epsilon^*_q$ immediately leads to
the mass of the nucleon in the medium in zeroth order as
\begin{equation}
E_N^{*0}=\sum_q~\epsilon^*_q
\label{eq:12}
\end{equation}

We next consider the spurious center-of-mass 
correction $\epsilon_{c.m.}$, the pionic correction $\delta M_{N}^\pi$ 
for restoration of chiral symmetry, and the 
short-distance one-gluon exchange contribution $(\Delta E_N)_g$ 
to the zeroth-order nucleon mass in the medium.
The center-of-mass correction $\epsilon_{c.m.}$ and the pionic corrections 
$\delta M_{N}^\pi$ in the present model are found, respectively, as \cite{rnm}
\begin{equation}
\epsilon_{c.m.}=\frac{(77\epsilon_u^{\prime}+31m_u^{\prime})}
{3(3\epsilon_u^{\prime}+m_u^{\prime})^2r_{0u}^2} \, 
\end{equation}
and
\begin{equation}
\delta M_{N}^\pi=- \frac{171}{25}I_{\pi}f_{NN\pi}^2.
\label{pion-corr}
\end{equation}
Here,
\begin{equation}
I_{\pi}=\frac{1}{\pi{m_{\pi}}^2}\int_{0}^{\infty}dk 
\frac{k^4u^2(k)}{w_k^2},
\end{equation} 
with the axial vector nucleon form factor given as 
\begin{equation}
u(k)=\Big[1-\frac{3}{2} \frac{k^2}{{\lambda}_q(5\epsilon_q^{\prime}+
7m_q^{\prime})}\Big]e^{-k^2r_0^2/4} \ .
\end{equation}
The pseudovector nucleon pion coupling constant $f_{NN{\pi}}$ can be obtained 
from the familiar Goldberg Triemann relation by using the axial-vector coupling-constant value $g_A$ in the model, 
as discussed in Ref. \cite{rnm}.

The color-electric and color-magnetic contributions to the gluonic correction 
which arises due to one-gluon exchange at short distances are given as:
\begin{equation}
(\Delta E_N)_g^E={\alpha_c}(b_{uu}I_{uu}^E+b_{us}I_{us}^E+b_{ss}I_{ss}^E) \ ,   
\label{enge}
\end{equation}
and due to color-magnetic contributions, as
\begin{equation}
(\Delta E_N)_g^M={\alpha_c}(a_{uu}I_{uu}^M+a_{us}I_{us}^M+a_{ss}I_{ss}^M) \ ,  
\label{engm}
\end{equation}
where $a_{ij}$ and $b_{ij}$ are the numerical coefficients depending on each 
baryon.
The color-electric contributions to the correction of baryon masses due to one gluon  
exchange are calculated in a field-theoretic manner \cite{rnm}. It can be found that 
the numerical coefficient for color-electric contributions such as 
$b_{uu}, b_{us}$, and $b_{ss}$ comes out to be zero. 
From calculations we have $a_{uu}=-3$ and  
$a_{us}=a_{ss}=b_{uu}=b_{us}=b_{ss}=0$ for the nucleons.
The quantities $ I_{ij}^{E} $ and $ I_{ij}^{M} $ are given in the following equation
\begin{eqnarray}
I_{ij}^{E}=\frac{16}{3{\sqrt \pi}}\frac{1}{R_{ij}}\Bigl[1-
\frac{\alpha_i+\alpha_j}{R_{ij}^2}+\frac{3\alpha_i\alpha_j}{R_{ij}^4}
\Bigl],
\nonumber\\
I_{ij}^{M}=\frac{256}{9{\sqrt \pi}}\frac{1}{R_{ij}^3}\frac{1}{(3\epsilon_i^{'}
+m_{i}^{'})}\frac{1}{(3\epsilon_j^{'}+m_{j}^{'})} \ ,
\end{eqnarray}
where 
\begin{eqnarray}
R_{ij}^{2}&=&3\Bigl[\frac{1}{({\epsilon_i^{'}}^2-{m_i^{'}}^2)}+
\frac{1}{({\epsilon_j^{'}}^2-{m_j^{'}}^2)}\Bigl]
\nonumber\\
\alpha_i&=&\frac{1}{ (\epsilon_i^{'}+m_i^{'})(3\epsilon_i^{'}+m_{i}^{'})} \ .
\end{eqnarray} 
In the calculation we have taken $\alpha_c= 0.58$ as the strong-coupling
constant in QCD at the nucleon scale \cite{barik}. The color-electric 
contribution is zero here, and the gluonic corrections to the mass of the 
nucleon are due to color-magnetic contributions only. 

Finally, treating all these corrections independently, the 
mass of the nucleon in the medium becomes 
\begin{equation}
M_N^*=E_N^{*0}-\epsilon_{c.m.}+\delta M_N^\pi+(\Delta E_N)^E_g+
(\Delta E_N)^M_g.
\label{mass}
\end{equation}
The total energy density and pressure at a 
particular baryon density for the nuclear matter becomes
\begin{eqnarray}
\label{engd}
{\cal E} &=&\frac{1}{2}m_\sigma^2 \sigma_0^2+\frac{1}{2}m_\omega^2 \omega^2_0
+\frac{1}{2}m_\rho^2 b^2_{03} \nonumber\\ &+&
\frac{\gamma}{(2\pi)^3}\sum_{N=p,n}\int ^{k_{f,N}} d^3 k \sqrt{k^2+{M_N^*}^2},\\
P&=&-~\frac{1}{2}m_\sigma^2 \sigma_0^2+\frac{1}{2}m_\omega^2 \omega^2_0+
\frac{1}{2}m_\rho^2 b_{03}^2\nonumber\\
&+&\frac{\gamma}{3(2\pi)^3}\sum_{N=p,n}\int ^{k_{f,N}} \frac{k^2~ d^3 k}
{\sqrt{k^2+{M_N^*}^2}},
\end{eqnarray}
where $\gamma=2$ is the spin degeneracy factor for nuclear matter. 
The nucleon density becomes
\begin{equation}
\rho_N= \frac{\gamma}{(2\pi)^3}\int_0^{k_{f,N}} d^3 k
=\frac{\gamma k_{f,N}^3}{6\pi^2}~~~\mbox{where} ~~N=p,n.
\end{equation}
Therefore, the total baryon density becomes $\rho_B=\rho_p+\rho_n$
and the (third component of) isospin density $\rho_3=\rho_p-\rho_n $.
The proton fraction, $y_p$ is defined as 
\begin{equation}
y_p=\frac{\rho_p}{\rho_B}
\end{equation}
where $\rho_p$ and $\rho_n$ are the proton and neutron densities.

The vector mean-fields $\omega_0$ and $b_{03}$ are determined through
\begin{equation}
\omega_0=\frac{g_\omega}{m_\omega^2} \rho_B~~~~~~~~~~
b_{03}=\frac{g_\rho}{2m_\rho^2} \rho_3,
\label{omg}
\end{equation}
where $g_\omega=3 g_\omega^q$ and $g_\rho= g_\rho^q$.
Finally, the scalar mean-field $\sigma_0$ is fixed by
\begin{equation}
\frac{\partial {\cal E }}{\partial \sigma_0}=0.
\label{sig}
\end{equation}
The iso-scalar scalar and iso-scalar vector couplings $g_\sigma^q$ and $g_\omega$
are fit to the saturation density and binding energy for nuclear
matter. The isovector vector coupling $g_\rho$ is set by fixing 
the symmetry energy. 
For a given baryon density, $\omega_0$, $b_{03}$, and $\sigma_0$ are
calculated from Eqs. \eqref{omg} and \eqref{sig}, respectively.

\section{The symmetry energy}
We may define the neutron-excess parameter $t=
\frac{\rho_n-\rho_p}{\rho_n+\rho_p}=(1-2y_p)$ so that the nuclear 
symmetry energy ${\cal E}_{sym}$ can be obtained as the difference between 
the total energy per baryon ${\cal E}/\rho_B=E(\rho_B,t)$ of pure neutron 
matter and that of isopspin-symmetric matter at baryon density 
$\rho_B$. Here we consider the nuclear matter consisting of protons and neutrons 
only with $y_p$  as the proton fraction. An expansion of the total energy per 
baryon, $E(\rho_B,t)$, with respect to the neutron-excess parameter, becomes 
\cite{steiner}
\begin{equation}
E(\rho_B,t)=E(\rho_B,0)+tE_1(\rho_B)+\frac{t^2}{2!}E_2(\rho_B)+
\frac{t^3}{3!}E_3(\rho_B)+\cdots,
\label{esymt}
\end{equation}
where $E_1, E_2, E_3, \cdots$, etc. are the first-, second- and third- order 
derivatives with respect to $t$ in a Taylor's expansion. However, neglecting Coulomb contributions near the isospin symmetry of QCD, demands the total energy 
of pure neutron matter to be same as that of pure proton matter, 
for which the odd powers in $t$ are to be forbidden in the above 
expansion. Again 
for densities near or below the saturation density ($\rho_B=\rho_0$), truncation 
of this expansion to quadratic terms in $t$ is considered to be a good 
approximation. In view of that, the coefficient of the quadratic term in $t$ 
can be identified as the symmetry energy
\begin{equation}
{\cal E}_{sym}(\rho_B)=\frac{1}{2}\left[\dfrac{\partial^2E(\rho_B)}
{\partial {t^2}}\right]_{t=0}=\frac{k_{f,N}^2}{6E_{f,N}^*}
+\frac{g_\rho^2} {8m_\rho^{2}}\rho_B,
\label{engs}
\end{equation}
where $E_{f,N}^*=(k_{f,N}^2+M_N^{*2})^{1/2}$.

This may be considered to be a good approximation even for small proton 
fraction $y_p$, which can be valid for finite nuclei. But for
nuclear matter at densities in excess of the saturation density $\rho_0$, 
effects of higher order than  quadratic in the expansion may be 
important. Therefore, in order to study the density dependence of 
${\cal E}_{sym}(\rho_B)$, one may
expand this as a function of $\rho_B$ around saturation density $\rho_0$ in
terms of a parameter $x=\frac{(\rho_B-\rho_0)}{3\rho_0}$ to obtain
\begin{equation}
{\cal E}_{sym}(\rho_B)=J+xL^0+\frac{x^2}{2!}K_{sym}^0+
\frac{x^3}{3!}Q_{sym}^0+\cdots,
\label{engs1}
\end{equation}
so as to consider the symmetry-energy parameters as follows: 
\begin{eqnarray}
J&=&{\cal E}_{sym}(\rho_0)\nonumber\\
L^0&=&3\rho_0\dfrac{\partial {\cal E}_{sym}(\rho_B)}{\partial \rho_B}
\Bigg|_{\rho_B=\rho_0}~{\mbox(Slope ~ of ~ {\cal E}_{sym})}\nonumber\\
K_{sym}^0&=&9\rho_0^2\dfrac{\partial^2 {\cal E}_{sym}(\rho_B)}{\partial 
\rho_B^2} \Bigg|_{\rho_B=\rho_0}{\mbox(Curvature ~ of ~ {\cal E}_{sym})}\nonumber\\
Q_{sym}^0&=&27\rho_0^3\dfrac{\partial^3 {\cal E}_{sym}(\rho_B)}{\partial 
\rho_B^3} \Bigg|_{\rho_B=\rho_0}{\mbox(Skewness ~ of ~ {\cal E}_{sym})}\nonumber\\
\label{cesym}
\end{eqnarray}
Apart from the quantities in Eq. \eqref{cesym}, the following quantities 
calculated from pressure $P$ and energy density ${\cal E}$ for the 
consideration of constraints and correlations studies are
\begin{eqnarray}
K_0&=&9\left[\frac{dP}{d\rho_B}\right]_{\rho_B=\rho_0,y_p=1/2}
~{\mbox(Compressibility)}\nonumber \\
Q_0&=&27\rho_0^3\dfrac{\partial^3 {\cal E}/\rho_B}{\partial 
\rho_B^3} \Bigg|_{\rho_B=\rho_0,y_p=1/2}{\mbox(Skewness ~ coefficient)}\nonumber\\
\label{ceng}
\end{eqnarray}
and the volume part of the iso-spin incompressibility
\begin{equation}
K_{{\tau},v}=K_{sym}^0-6L^0-\frac{Q_0}{K_0}L^0.
\end{equation}
We have assumed $K_{{\tau},v}=K_{\tau}$ since the 
volume term is dominant \cite{dutraetal}. 
These parameters characterize the density dependence of nuclear symmetry 
energy around normal nuclear matter density and thus provide important 
information on the behavior of nuclear symmetry energy at both high and 
low densities.  Also, the curvature parameter $K_{sym}^0$ 
distinguishes the different parametrizations. 
A more significant measurement would be the evaluation of 
the shift of the incompressibility with asymmetry, which is given by
\begin{equation}
K_{asy}=K_{sym}^0-6L^0
\end{equation}
because this value can be correlated to experimental observations of the 
giant monopole resonance (GMR) of neutron-rich nuclei. Recent observations 
of the GMR \cite{tili} on even-$A$ Sn isotopes give a quite stringent value of 
$K_{asy}=-550 \pm 100$ MeV. In the present model we determine this value 
for three quark masses of $300$, $40$, and $5$ MeV and observe that 
they are consistent with the GMR measurements.

The compressibility $K_0$ at saturation density can be determined analytically, from Eq. \eqref{ceng}:
\begin{eqnarray}
K_0&=&9\left(\frac{g_\omega}{m_\omega}\right)^2\rho_0+\frac{3k_{f,N}^2}{E_{f,N}^*}
+\frac{3k_{f,N}M_N^*}{E_{f,N}^*}\cdot\frac{dM_N^*}{dk_{f,N}}.
\end{eqnarray}
The study of the correlation between symmetry energy and its slope can be 
performed analytically. For this purpose we use the Eqs. \eqref{engd} and 
\eqref{engs} to find ${\cal E}_{sym}$. 
In this model, we get the closed-form expression
%\begin{eqnarray}
%L&=&3\rho_0\dfrac{\pi^2}{2k_{f,N}^2}\dfrac{\partial}{\partial k_{f,N}}{\cal E}_{sym}(\rho_0)\nonumber\\
%&=&k_{f,N}\dfrac{\partial}{\partial k_{f,N}}\left[\dfrac{g_\rho^2}{m_\rho^2}\dfrac{k_{f,N}^3}{12\pi^2}
%+\dfrac{k_{f,N}^2}{6E_{f,N}^*}\right]\nonumber\\
%&=&\dfrac{g_\rho^2}{m_\rho^2}\dfrac{k_{f,N}^3}{4\pi^2}+
%\dfrac{k_{f,N}^2}{3E_{f,N}^*}-\dfrac{k_{f,N}^4}{6E_{f,N}^{*^3}}\nonumber \\
%&-&\dfrac{k_{f,N}^2}{18E_{f,N}^{*^2}}\left(\dfrac{3k_{f,N}M_N^*}{E_{f,N}^*}
%\dfrac{dM_N^*}{dk_{f,N}}
%\right)
%\end{eqnarray}
%Converting $k_{f,N}$ into $\rho_{0}$ and with a little algebra we obtain
%\begin{eqnarray}
%L&=&3{\cal E}_{sym}+\frac{1}{2}\left(\frac{3\pi^2}{2}\right)^{2/3}
%\frac{1}{E_{f,N}^*} \nonumber\\
%&\times&\left[\left(\frac{g_\omega}{m_\omega}\right)^2\frac{\rho_0^{5/3}}
%{E_{f,N}^*}- \frac{\rho_0^{2/3}K_0}{9E_{f,N}^*}-\frac{\rho_0^{2/3}}{3} \right] \nonumber\\
%&=&3{\cal E}_{sym}+f(M_N^*,\rho_0,B_0,K_0)
%\end{eqnarray}
\begin{equation}
L^0=3J+f(M_N^*,\rho_0,B_0,K_0),
\end{equation}

where 
\begin{eqnarray}
f(M_N^*,\rho_0,&&B_0,K_0)=\frac{1}{2}\left(\frac{3\pi^2}{2}\right)^{2/3}
\frac{1}{E_{f,N}^*}\times\nonumber\\
&&\left[\left(\frac{g_\omega}
{m_\omega}\right)^2\frac{\rho_0^{5/3}}{E_{f,N}^*}-\frac{\rho_0^{2/3}K_0}
{9E_{f,N}^*}-\frac{\rho_0^{2/3}}{3}\right].
\end{eqnarray}
The correlation function $f(M_N^*,\rho_0,B_0,K_0)$ exhibits the dependence on 
the different bulk parameters $M_N^*, \rho_0, B_0$, and $K_0$.

\section{Stability Conditions}
Nuclear forces have an attractive long-range part and a repulsive hard core 
similar to a Van der Waals fluid. It is expected to present a  
liquid and a gas phase characterized by the respective densities. Nucleons 
can be either protons or neutrons. Such a two-component system undergoes 
liquid-gas phase transition. The asymmetric nuclear matter (ANM) shows 
two types of instabilities \cite{chomaz}: a mechanical instability conserving the proton 
concentration and a chemical instability occurring at constant density.

We consider asymmetric nuclear matter characterized by proton and neutron
densities $\rho_N=\rho_p,~\rho_n$ and transform these into a set of two mutually
commuting charges $\rho_i=\rho_B,~\rho_3$ \cite{unique}.
In infinite matter the extensivity of free-energy implies that it can be 
reduced to a free energy density: ${\cal F} (T,\rho_i)={\cal E}-TS$ which
at $T=0$ reduces to energy density ${\cal E}$ only.
Since, we deal with a two-component nuclear medium, spinodal instabilities are
intimately related to phase equilibria and phase transitions. Although it
consists of unstable states, the spinodal region of the phase diagram can
be addressed by standard thermodynamics.

The condition for stability implies that the free energy density is a 
convex function of the densities $\rho_i$. A local necessary condition is 
the positivity of the curvature matrix:
\begin{equation}
{\cal F}_{ij}=\left(\frac{\partial^2{\cal F}}{\partial \rho_i\partial\rho_j}
\right)_T\equiv\left(\frac{\partial \mu_i}{\partial\rho_{j}}\right)_T.
\label{stability}
\end{equation}
Here we used $\left.\frac{\partial{\cal F}}{\partial \rho_i}
\right|_{T,\rho_{j\ne i}}=\mu_i$,
where the effective chemical potentials in the present context are given by
\begin{eqnarray}
\mu_p&=&\sqrt{k_{f,p}^2+{M^*_p}^2}+V_\omega+\frac{1}{2} V_\rho,\nonumber\\
\mu_n&=&\sqrt{k_{f,n}^2+{M^*_n}^2}+V_\omega-\frac{1}{2} V_\rho.
\end{eqnarray}
Since we consider a two-fluid system, $\left[{\cal F}_{ij}\right]$ is a 
$2\times 2$ 
symmetric matrix with two real eigenvalues $ \lambda_{\pm}$ \cite{ms}. The two 
eigenvalues are given by,
\begin{equation}
\lambda_{\pm}=\frac{1}{2}\left(\mbox{Tr}({\cal F})\pm\sqrt{\mbox[{Tr}({\cal
F})]^2-4\mbox{Det}({\cal F})}\right),
\end{equation}
and the eigenvectors $\boldsymbol{\delta\rho_\pm}$ by
\begin{equation}
\frac{\delta\rho^\pm_i}{\delta\rho^\pm_j}=\frac{\lambda_\pm-{\cal F}_{jj}}
{{\cal F}_{ji}}, \quad i,j=p,n.
\end{equation}
The largest eigenvalue is always positive whereas the
other can take on negative value. We are interested in the 
latter, because it defines the spinodal surface,
which is determined by the values of $T, \rho,$, and $y_p$. 
The smallest eigenvalue  of ${\cal F}_{ij}$ becomes negative. 
The associated eigenvector defines the instability direction of the system, 
in isospin space. 

\section{Results and Discussion}
We set the model parameters $(a,V_0)$ by fitting the nucleon mass $M_N=939$ MeV 
and charge radius of the proton $\langle r_N\rangle=0.87$ fm in free space. 
Taking standard values for the meson masses; namely, $m_\sigma=550$ MeV, 
$m_\omega=783$ MeV and $m_\rho=763$ MeV and fitting the quark-meson coupling 
constants self-consistently, we obtain the correct saturation 
properties of nuclear matter binding energy, $E_{B.E.}\equiv B_0={\cal E}/\rho_B-M_N=-15.7$ 
MeV, pressure, $P=0$, and 
symmetry energy $J=32.0$ MeV at $\rho_B=\rho_0=0.15$ fm$^{-3}$.
The values of $g_\sigma^q$, $g_\omega$, and $g_\rho$ 
obtained this way and the values of the model parameters at quark 
masses $5$, $40$, and $300$ MeV are given in Table \ref{table2}.
\begin{table}[t]
\centering
\renewcommand{\arraystretch}{1.4}
\setlength\tabcolsep{3pt}
\begin{tabular}{|c|c|c|c|c|c|}
\hline
$m_q$ (MeV)& $g^q_\sigma$& $g_\omega$& $g_\rho$& $a($fm$^{-3})$& $V_0($MeV$)$\\
\hline
5     &6.44071  &2.39398 &9.04862 &0.978629 &111.265238   \\
\hline
40     &5.46761  &3.96975 &8.99036 &0.892380 &100.187229   \\
\hline
300   &4.07565  &9.09078  &8.51458 &0.534296 &-62.257187  \\
\hline
\end{tabular}
\caption{\label{table2}Parameters for nuclear matter. They are determined 
from the binding energy per nucleon, $E_{B.E}=B_0 \equiv{\cal E} /\rho_B - M_N 
= -15.7$~MeV and pressure, $P=0$ at saturation density 
$\rho_B=\rho_0=0.15$~fm$^{-3}$.}
\end{table}
%%%%%%%%%%%%%%%%%%%
\begin{table*}[t]
  \centering
\caption{Nuclear matter properties of the models used in the present work. 
The quantities presented are at saturation density.}
  \begin{tabular}{lclccccccccc}
\hline
Model & $B_0$ & $~\rho_0$ & $M^*/M$ &  $K_0$ & $J$ & $L^0$
& $K_{\rm{sym}}^0$ & $K_{\rm{asy}}$ & $Q_0$ &$~K_\tau$  \\
          & (MeV)&(fm$^{-3}$) & & (MeV) & (MeV)& (MeV)& (MeV) & (MeV) & (MeV) &(MeV)\\
\hline
MQMC (5 MeV)                      &-15.7 &0.151 &  0.93 &159 & 32.0  & 84.7  & -27.7 & -535.9& 103.2 &-590.8 \\
MQMC (40 MeV)                     &-15.7 &0.151 &  0.91 &208 & 32.0  & 84.9  & -28.4 & -537.6& 94.2  &-575.9\\
MQMC (300 MeV)                    &-15.7 &0.151 &  0.76 &349 & 32.0  & 89.1  & -14.5 & -549.0& -15.6 &-545.1\\
DD~\cite{typel}                   &-16.0 &0.149 &  0.56 &239 & 31.6  & 55.9  & -95.3 & -431.1& 576.8 &-462.57\\
QMC~\cite{alex,alextemp}          &-15.7 &0.150 &  0.77 &291 & 33.7  & 93.5  & -10.0 & -570.8& 29.4  &-580.24 \\
FSUGold~\cite{todd}               &-16.2 &0.148 &  0.61 &229 & 32.6  & 60.4  & -51.4 & -414.0& 425.7 &-276.07\\
BKA24~\cite{bka24}                &-15.9 &0.147 &  0.60 &227 & 34.2  & 84.8  & -14.9 & -523.7& 112.4 &-421.55\\
BSR12~\cite{bsr12}                &-16.1 &0.147 &  0.61 &232 & 34.0  & 77.9  & -44.2 & -511.6& 324.2 &-414.30\\
\hline
  \end{tabular}
  \label{tab:properties}
\end{table*}
%%%%%%%%%%%%%%%%%%%
\begin{figure*}
\includegraphics[width=8.cm,angle=0]{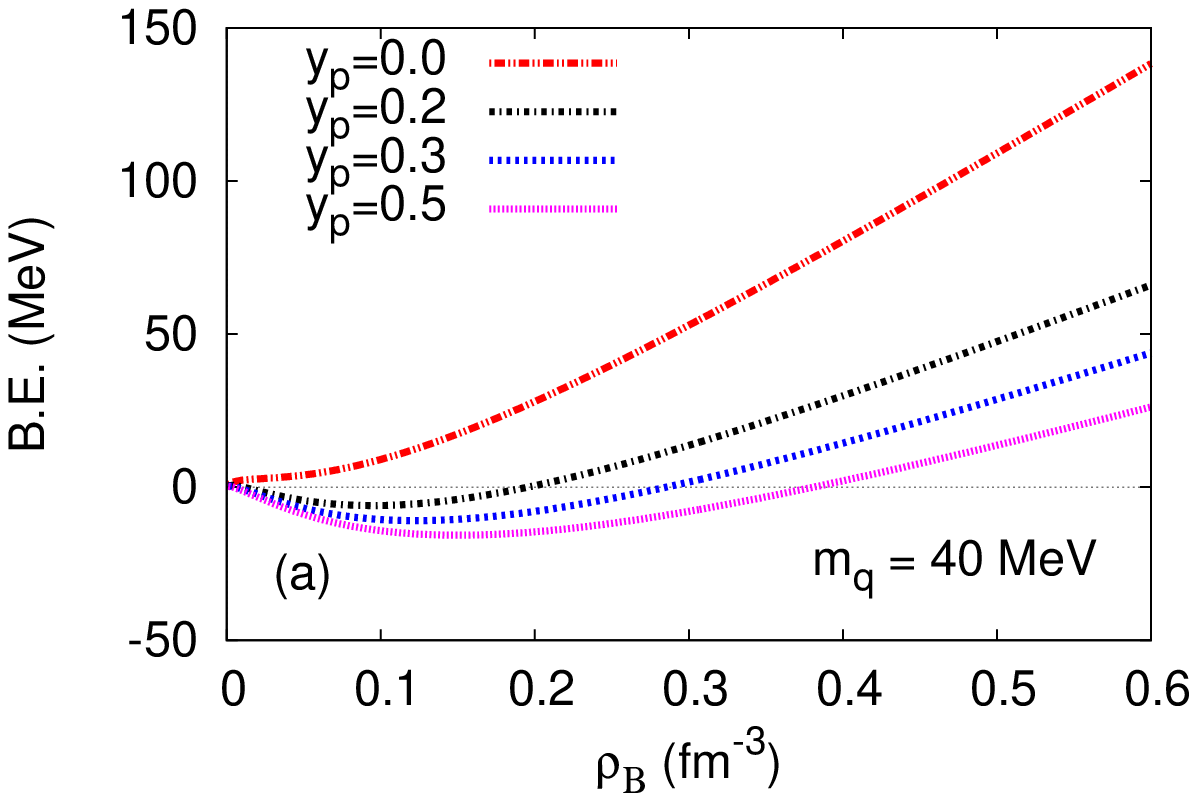}
\includegraphics[width=8.cm,angle=0]{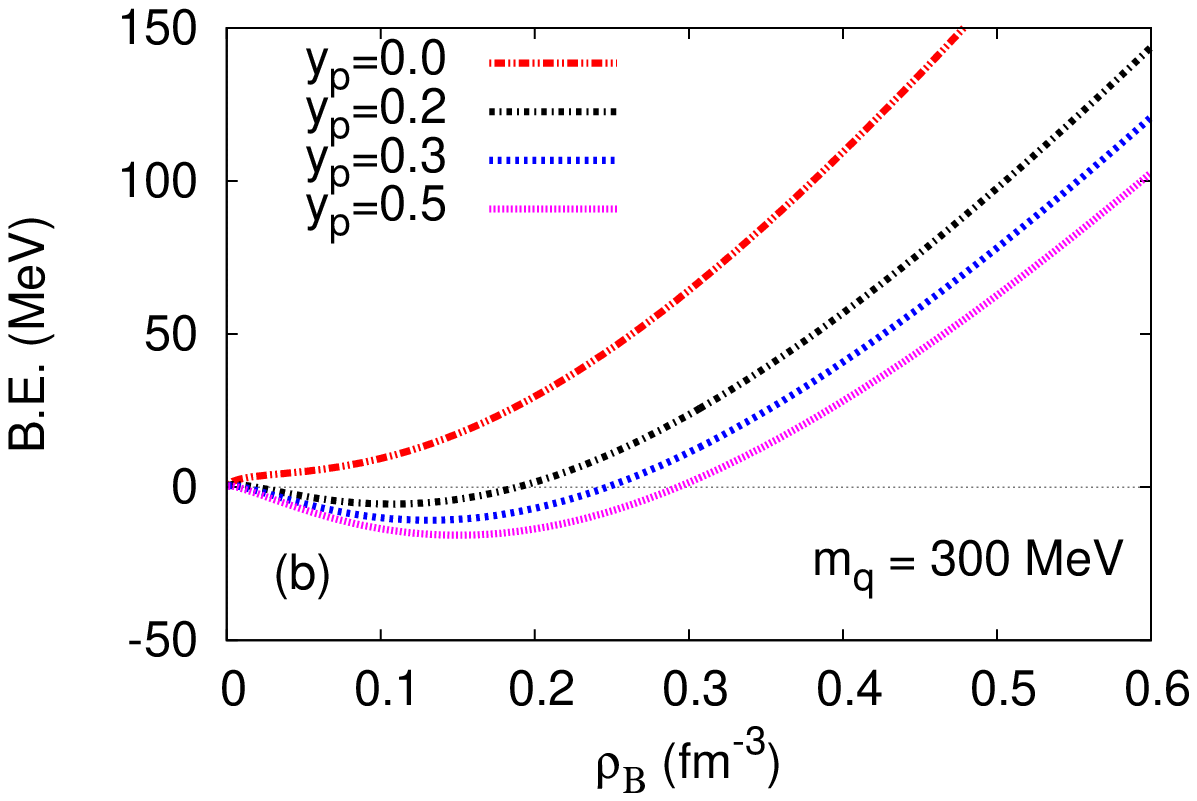}
\caption{(Color online) Nuclear matter binding energy as a function of density 
for (a) different $y_p$ values for quark mass $m_q=40$ MeV and (b) quark mass 
$m_q=300$ MeV.}
\label{fig1}
\end{figure*}
In Figs. \ref{fig1}(a) and \ref{fig1}(b), we plot the binding energy per 
nucleon for nuclear matter as a function of density corresponding to $m_q=40$ 
MeV and $m_q=300$ MeV, respectively, for different $y_p$ values. In Fig. \ref{fig1a}, we 
compare the variation of the binding energy per nucleon for quark mass $40$ and 
$300$ MeV with that of QMC and observe that,
for $300$ MeV, MQMC compares well with that of QMC. 
It is observed from Figs. \ref{fig1} and \ref{fig1a} that, at low quark
mass, the equation of state is softer. In Table \ref{tab:properties}, 
we compare the 
nuclear matter properties at saturation for quark masses $5$, $40$, and 
$300$ MeV, respectively, in the present model to QMC \cite{alex,alextemp}, 
and some of the approved models as suggested in Ref. \cite{dutraetal}.

The value of the compressibility 
$K_0$ is determined to be $159$, $208$, and $349$ MeV respectively, for 
quark masses $5$, $40$, and $300$ MeV. A recent calculation \cite{stonemos} has
predicted $K_0$ to be in the range $250~<~K_0~<~315$ from the experimental
GMR energies in even-even $^{112-124}$Sn and $^{106,100-116}$Cd. Furthermore, 
the value of the effective mass calculated in the present model at quark mass 
$300$ MeV is $0.76$ which compares well with the empirical value of the 
effective mass, which is $0.74$ \cite{mahaux}. 
\begin{figure}
\includegraphics[width=8.cm,angle=0]{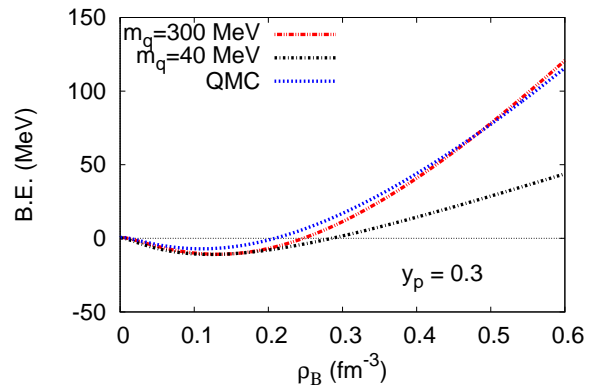}
\caption{(Color online) Nuclear matter binding energy as a function of 
density for quark mass $m_q=40$ MeV and quark mass $m_q=300$ MeV for $y_p=0.3$. 
A comparison is made between the MQMC for quark mass $40$ MeV and $300$ MeV 
with that of QMC.}
\label{fig1a}
\end{figure}

\begin{figure}
\includegraphics[width=8.cm,angle=0]{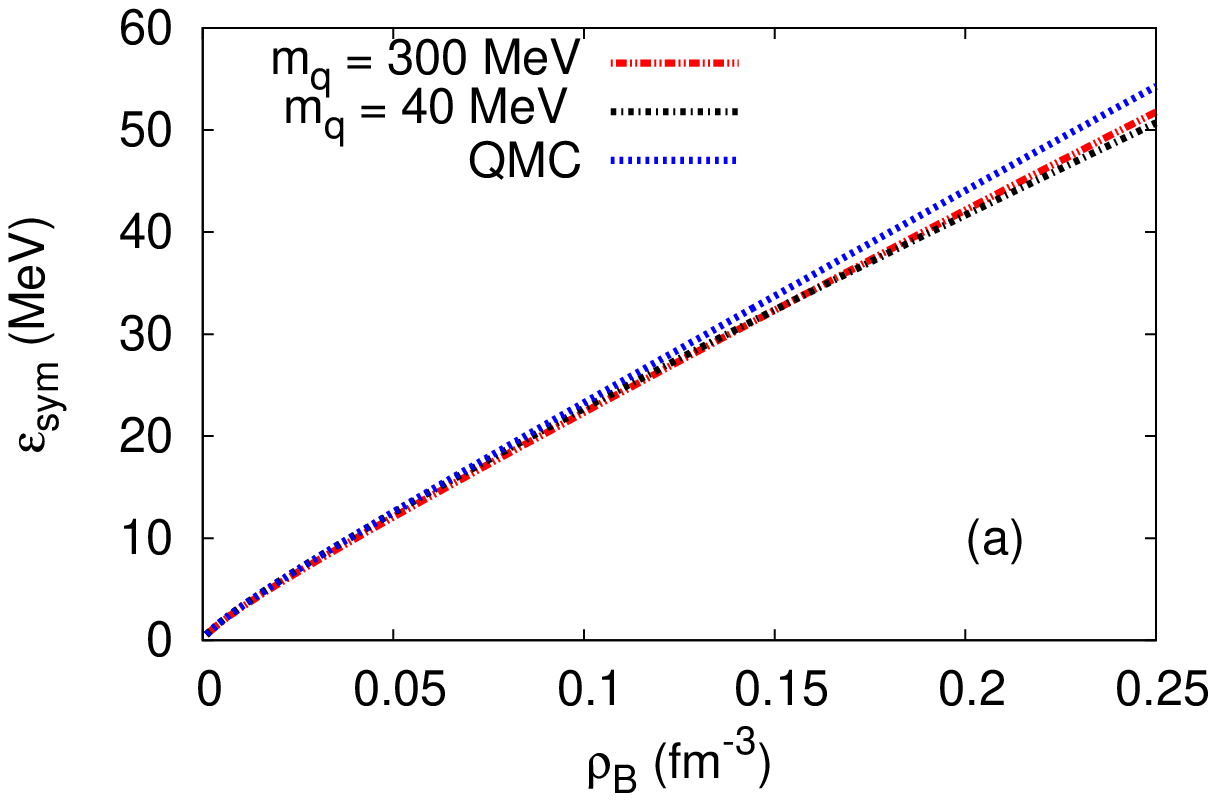}
\includegraphics[width=8.cm,angle=0]{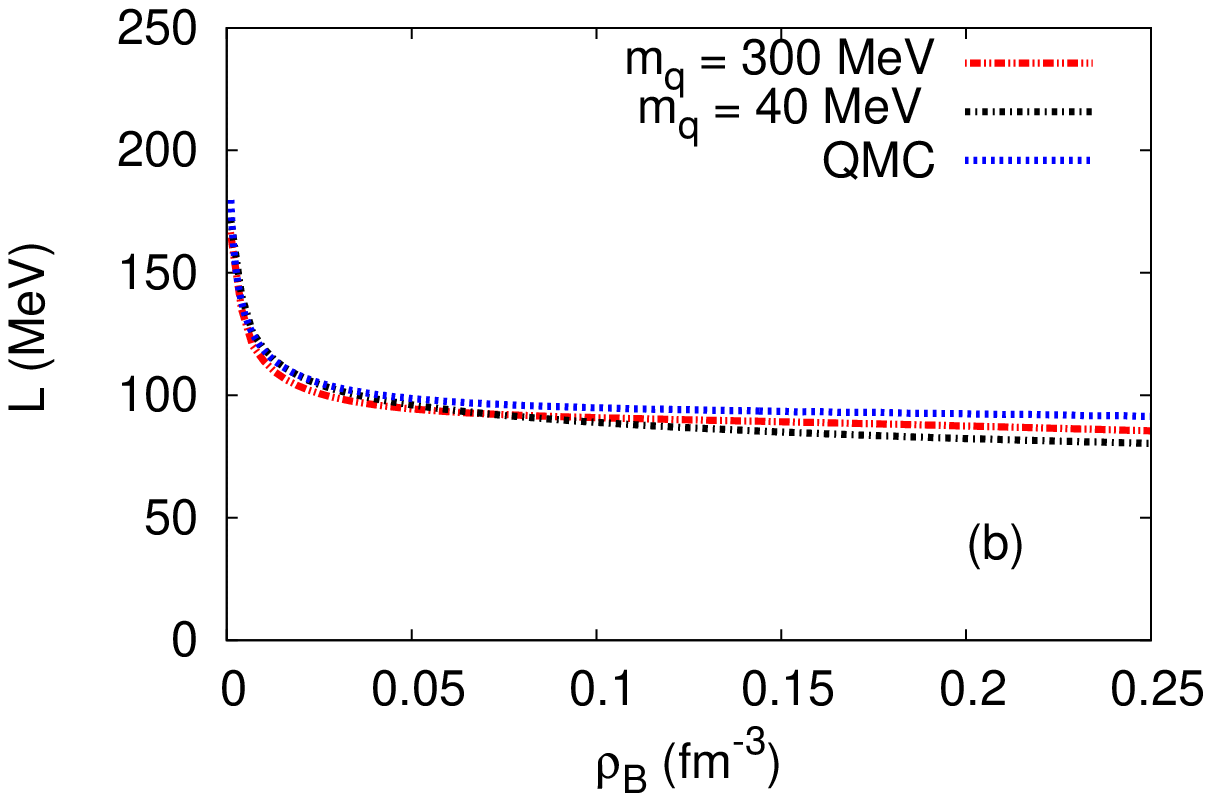}
\includegraphics[width=8.cm,angle=0]{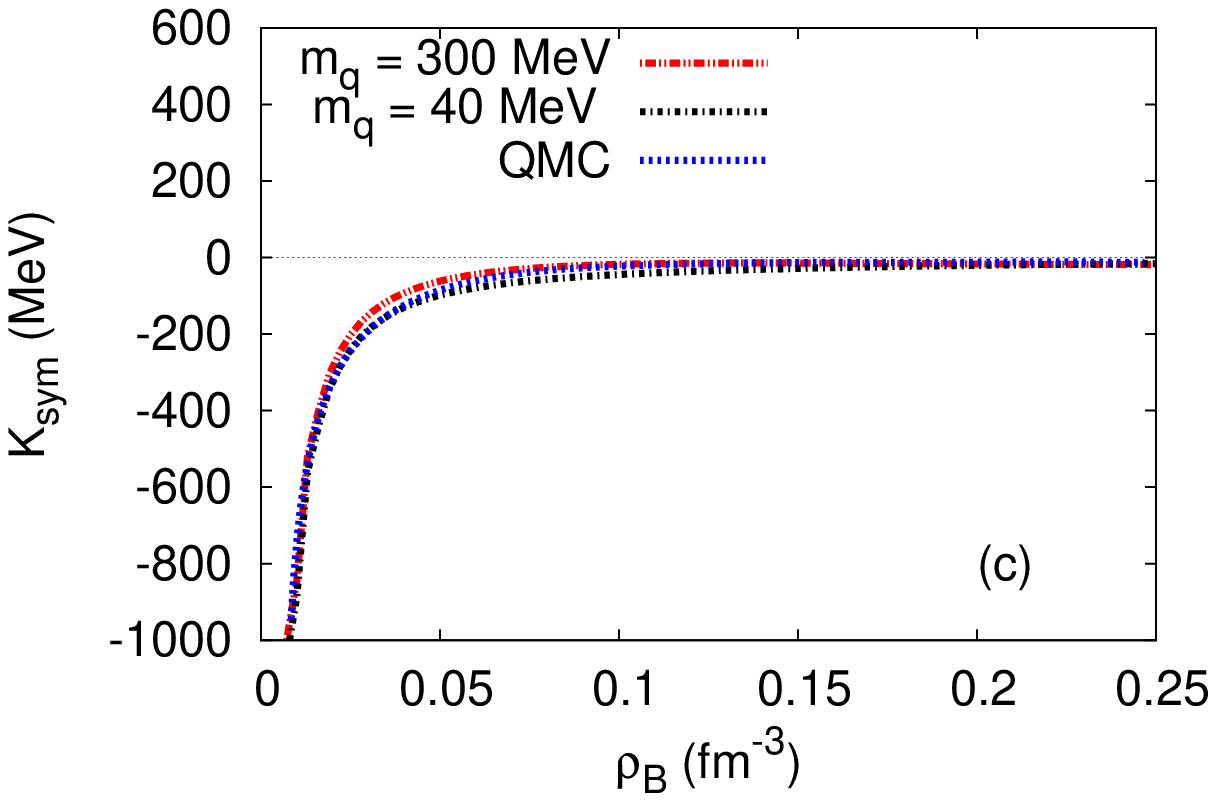}
\caption{(Color online) (a) Symmetry energy, (b) its slope parameter 
$L$, and (c) $K_{sym}$ (c) in the MQMC and QMC models as functions of 
baryon density $\rho_B$}.
\label{fig2}
\end{figure}

\begin{figure}
\vspace{-0.5in}
\includegraphics[width=8.cm,angle=0]{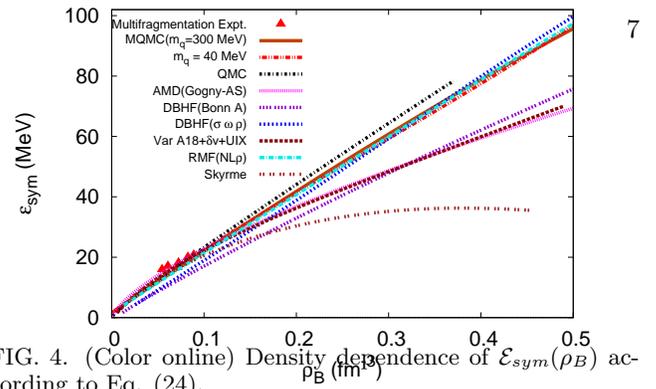}
\vspace{-0.5in}
\caption{(Color online) Density dependence of ${\cal E}_{sym}(\rho_B)$ 
according to Eq. \eqref{engs1}.} 
\label{fig3a}
\end{figure}

We compare the symmetry energy, its slope, and incompressibility from our model 
with the QMC \cite{alex,alextemp} results respectively in Figs. \ref{fig2}(a), 
 \ref{fig2}(b), and \ref{fig2}(c). 
We observe that the symmetry 
energy shows an extremely linear behavior. 
This is further justified from the plot for the slope parameter $L$. 
This is based on equation \eqref{engs} for ${\cal E}_{sym}$. However, 
if we consider terms higher than the quadratic one in defining the relation
in Eq. \eqref{esymt}, it would be more appropriate to use the expression as in 
Eq. \eqref{engs1} to show the density dependence of ${\cal E}_{sym}$
for higher nuclear densities $(\rho_B > \rho_0)$. This has been shown in Fig. 
\ref{fig3a} in comparison with several other models, as noted there, 
including QMC.
\subsection{Correlation between the symmetry energy and its slope}
\begin{figure}
\includegraphics[width=8.5cm,angle=0]{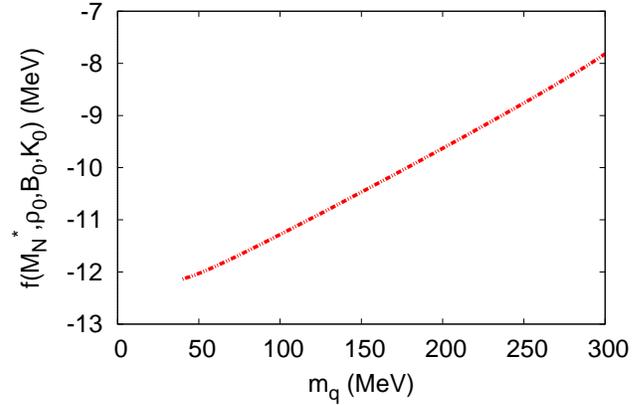}
\caption{\label{fig9}(Color online) Correlation function 
$f(M_N^*,\rho_0,B_0,K_0)$ at various quark masses.}
\end{figure}
We study the correlation function $f(M_N^*,\rho_0,B_0,K_0)$ with the 
variation of quark masses in Fig. \ref{fig9}. 
We observe that the function $f(M_N^*,\rho_0,B_0,K_0)$ increases with quark 
masses. 
\begin{figure}
\includegraphics[width=8.5cm,angle=0]{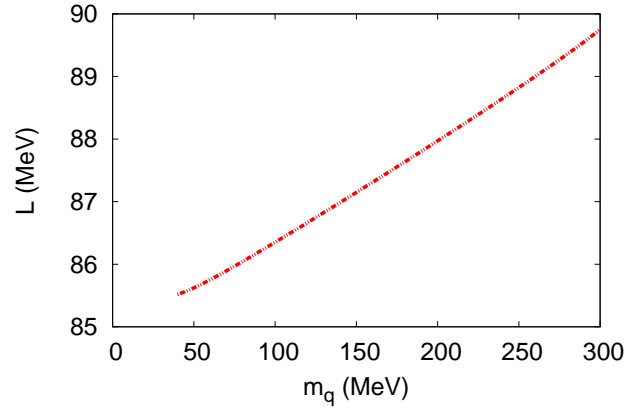}
\caption{\label{fig10}(Color online) The slope of symmetry energy, $L$ at
various quark mass.}
\end{figure}
The established value of binding energy $B_0$ and the saturation density $\rho_{0}$ in the nuclear 
mean-field models cannot be applied to incompressibility $K_0$ and effective mass 
$M_N^*$ since the latter are found as output in this model where the 
coupling constants are fixed in a self consistent manner by taking into 
consideration the binding energy and saturation density.
Therefore, we have taken the variation of $f$ with different quark masses 
at the same $B_0$ and $\rho_{0}$. We observe that, because there is only one isovector parameter $g_{\rho}^2$ 
in the expression for ${\cal E}_{sym}$ and $L$,  the variation is linear. Such linearity in the behavior 
was also observed in nonrelativistic models \cite{correl}. It indicates one of the 
limitations of the model parameters. We expect a nonlinear behavior 
between ${\cal E}_{sym}$ and $L$ for the models with more than one isovector 
parameter. In the Fig. \ref{fig10}, we have shown the  slope of symmetry 
energy, $L$ at various quark masses.  It is interesting to note 
that there is a linear relationship of the slope of the symmetry energy $L$ 
with quark mass. This is a direct 
consequence of the dependence of the symmetry energy on $g_\rho^{2}$.

\subsection{Instability}
\begin{figure}
\includegraphics[width=8.cm,angle=0]{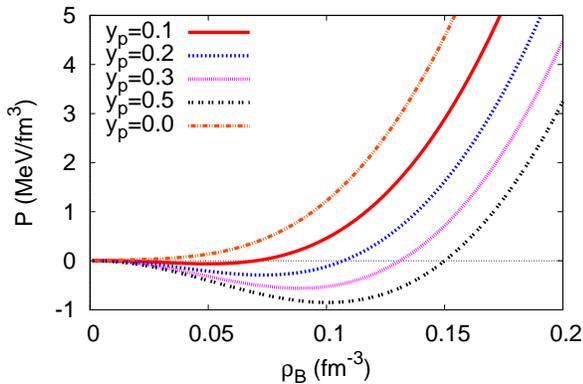}
\caption{(Color online) Pressure as a function of density at various isospin asymmetries.}
\label{fig3}
\end{figure}

We next study the mechanical instability and its 
dependence on the isospin asymmetry of the system by plotting the pressure as 
a function of density and the asymmetry parameter $y_p$. In Fig. \ref{fig3}, 
we show that the mechanical instability occurs in the region where the slope 
of the pressure with respect to density is negative. We observe that the 
mechanical-instability region shrinks when the isospin-asymmetry increases.
The system is stable under separation of two phases if the free energy of a 
single phase is lower than the free energy in all two-phase configurations. 

In the spinodal area we can get the signature of the mechanical 
instability by finding the velocity of sound determined in the model as
\begin{equation}
\beta^2=\dfrac{dP}{d\cal E }= 
\dfrac{dP}{d\rho_B}/{\dfrac{d\cal E}{d\rho_B}},
\end{equation}
where $\beta^2=v_s^2/c^2$, $v_s$ is the velocity of sound and $c$ is the 
speed of light.
\begin{figure}
\includegraphics[width=8.cm,angle=0]{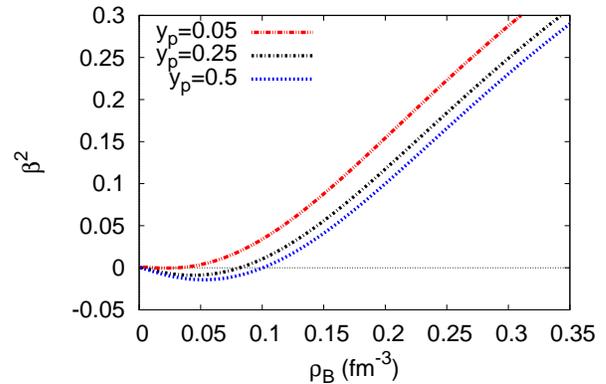}
\caption{\label{fig4}(Color online) Sound velocity as a function of density.}
\end{figure}
In Fig. \ref{fig4} we show the sound velocity as a function of density 
by changing the
asymmetry parameter. There is a reduction in the instability when we move away
from the symmetric nuclear matter. Moreover, the velocity becomes imaginary
when we enter into the spinodal area \cite{unique}.

\begin{figure}
\includegraphics[width=8.cm,angle=0]{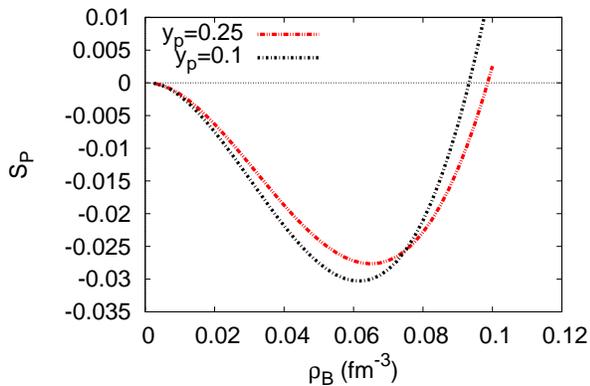}
\caption{\label{fig5}(Color online) The stability parameter 
$S_p$ [Eq. \eqref{stab}] 
as a function of $\rho_B$ for $y_p=0.25$ and $y_p=0.1$ in the instability sector.}
\end{figure}

The positivity of the local curvature matrix is equivalent to the condition 
that both the trace $Tr\left[{\cal F}_{ij}\right]=\lambda_++\lambda_-$ and the 
determinant $det\left[{\cal F}_{ij}\right]=\lambda_+\lambda_-$are positive. 
In the present model the  above condition is violated and the system is in 
the unstable region of a phase transition. Further it is to be pointed out that 
for a two component, n-p thermodynamical system, the stability parameter is 
given by the condition;
\begin{equation}
S_P=\left(\frac{\partial P}{\partial \rho_B}\right)_{T,y_p}\cdot
\left(\frac{\partial \mu_p}{\partial y_p}\right)_{T,P} > 0,
\label{stab}
\end{equation}
but in charge symmetric matter the isoscalar (total density) $\delta\rho_n+
\delta\rho_p$ and isovector (concentration) $\delta\rho_n-\delta\rho_p$ 
oscillations are not coupled and there are two separate conditions for 
instability \cite{asym}. These conditions are for mechanical instability
\begin{equation}
\left(\frac{\partial P}{\partial \rho_B}\right)_{T,y_p}\leq 0,
\label{pom1}
\end{equation}
and  for chemical instability
\begin{equation}
\left(\frac{\partial \mu_p}{\partial y_p}\right)_{T,P}\leq 0.
\label{pom2}
\end{equation}
In the ANM the isoscalar and isovector modes are coupled and the 
two separate inequalities do not select the nature of instability. Moreover, we 
observe in Fig. \ref{fig5}, a large difference in the behavior of the 
stability parameter $S_P$ in Eq. \eqref{stab} inside the instability 
region. For higher asymmetry, the range of the stability parameter is smaller 
than at lower asymmetries.
To understand this effect we follow the 
Landau-dispersion-relation approach for small-amplitude oscillations in 
Fermi liquids \cite{landau}. For a two component $(n,p)$ matter, the interaction is 
characterized by the Landau parameters $F^{q,q'}_0$ which is defined by the 
relationship
\begin{equation}
N_q(T)\frac{\partial \mu_q}{\partial \rho_{q'}} \equiv \delta_{q,q'}+F_0^{q,q'},
\end{equation}
where $q=(n,p)$ and $N_q(T)$ represents the single-particle level density at the Fermi 
energy. At zero temperature it has the simple form
\begin{equation}
N_q = \frac{k_{Fq}E^{*}_{Fq}}{\pi^2}.
\end{equation}
In the symmetric case ($F^{nn}_0=F^{pp}_0,F^{np}_0=F^{pn}_0 $), the Eqs. \eqref{pom1} and \eqref{pom2}
correspond to the two Pomeranchuk instability conditions
\begin{eqnarray}
F^{s}_0 &=& F^{nn}_0+F^{np}_0 < -1~~~~~~{\mbox mechanical} \nonumber\\
F^{a}_0 &=& F^{nn}_0-F^{np}_0 > -1~~~~~~{\mbox chemical}
\end{eqnarray}
The dispersion relations $F^{s}_0$ give the properties
of density (isoscalar) modes and $F^{a}_0$ gives the concentration
(isovector) modes.

\begin{figure}
\includegraphics[width=8.cm,angle=0]{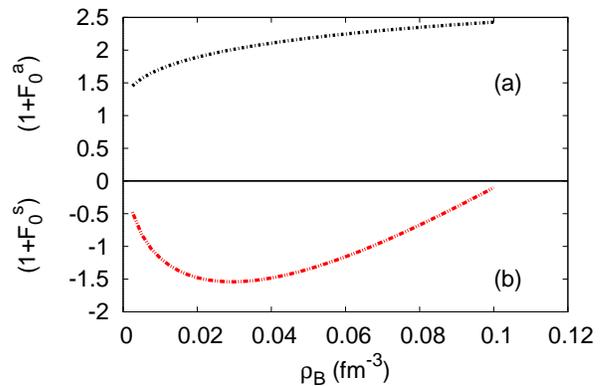}
\caption{\label{fig6}(Color online) Behavior of generalized Landau parameters 
(a) $1+F^{a}_0$ (b) $1+F^{s}_0$ with respect to baryon density $\rho_B$ 
in the instability sector for $y_p=0.25$.}
\end{figure}

In the unstable region of dilute asymmetric nuclear matter 
we have isoscalar-like unstable modes, hence $1+F^{s}_0<0$, while the 
combination $1+F^{a}_0>0$. In the Fig. \ref{fig6} we plot the values 
obtained from the calculation of these two quantities in the unstable 
region at zero temperature for $y_p=0.25$. An important observation we 
make from the comparison of Figs. \ref{fig5} and \ref{fig6} is 
the shift in the maximum instability density region. 
In Fig. \ref{fig5} the largest instability (the most negative value) is at 
$\rho_B=0.065 fm^{-3}$. However, the most negative Pomeranchuk condition 
$1+F_0^s$, which corresponds to the fastest unstable mode, is present 
in more dilute matter at $\rho_B=0.03 fm^{-3}$.

\begin{figure}
\includegraphics[width=8.cm,angle=0]{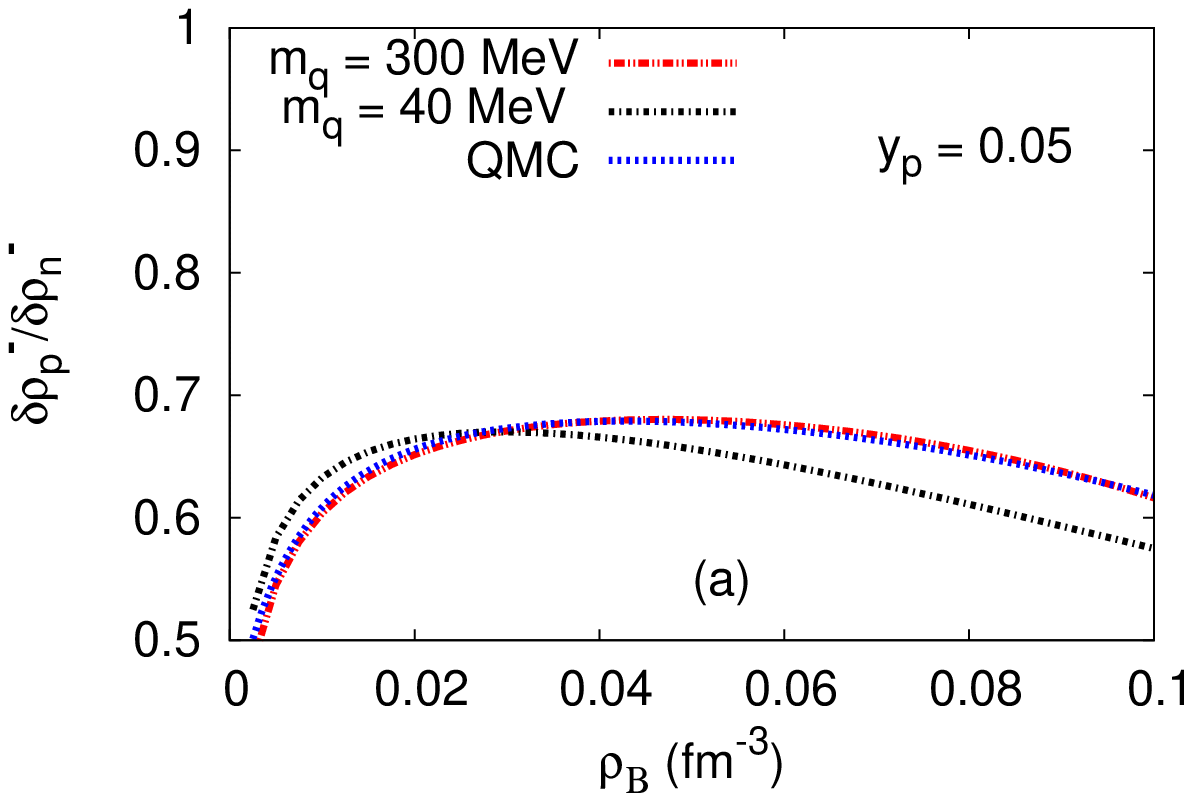}
%\caption{\label{fig3}Direction of instability.
%$m_q=40$ MeV and $m_q=300$ MeV.}
%\end{figure}
%\begin{figure}
\includegraphics[width=8.cm,angle=0]{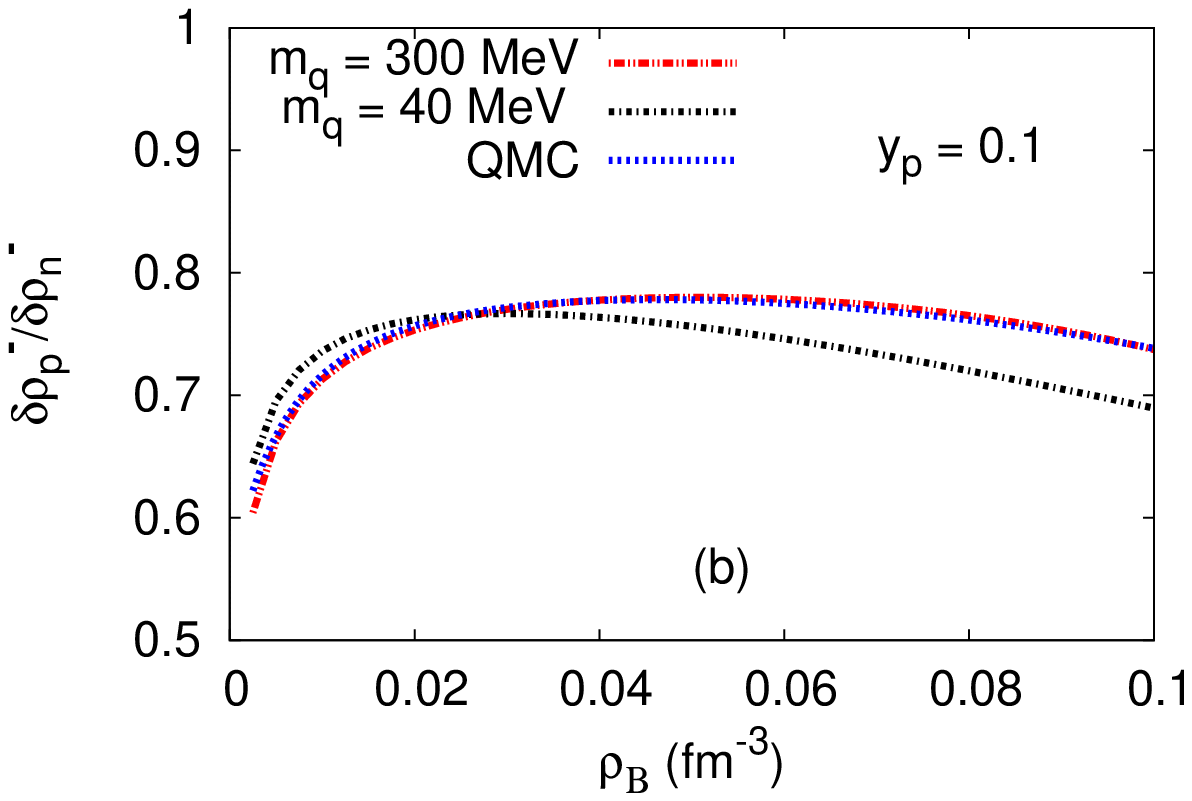}
%\caption{\label{fig3}Direction of instability.
%$m_q=40$ MeV and $m_q=300$ MeV.}
%\end{figure}
%\begin{figure}
\includegraphics[width=8.cm,angle=0]{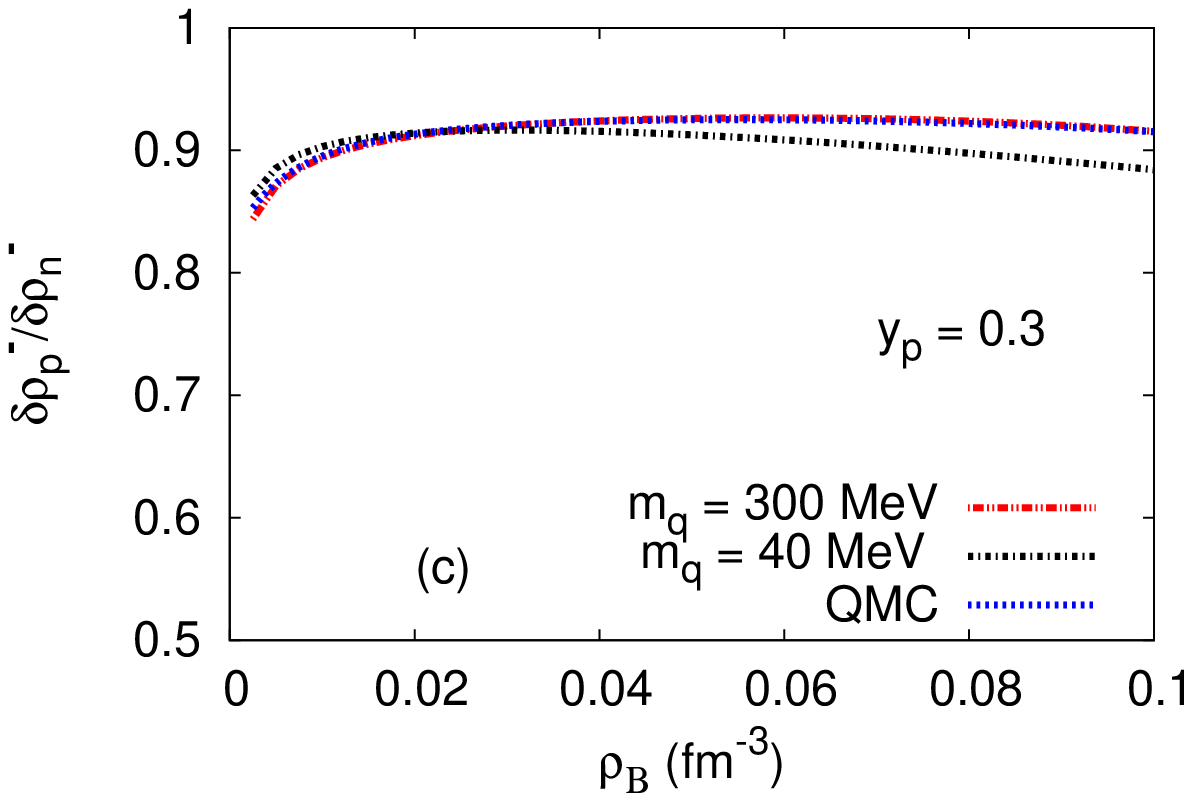}
\caption{\label{fig7}(Color online) Ratio of proton-neutron density fluctuation 
corresponding to the unstable mode showing the direction of instability for 
(a) $y_p=0.05$, (b) $y_p=0.1$, and (c) $y_p=0.3$.}
\end{figure}

In the following we study the direction of instability of the system. In Fig.
\ref{fig7}, we show the ratio of the proton versus neutron density fluctuations 
corresponding to the unstable mode 
which defines the direction of instability of the system. We plot the results 
for different proton fractions and observe that the instabilities tend to 
restore the isospin symmetry for the dense (liquid) phase leading to the 
fractionation of the ANM.

\begin{figure}
\includegraphics[width=8.cm,angle=0]{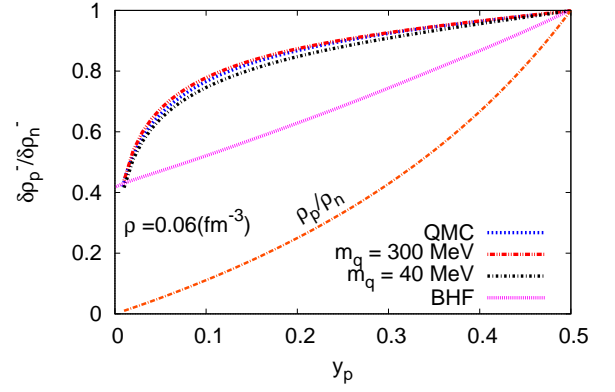}
\caption{\label{fig8}(Color online) Proton-neutron density fluctuation ratio
versus proton fraction $y_p$ for a fixed nuclear density.}
\end{figure}
Figure \ref{fig8} shows the proton-neutron density fluctuation ratio as a function 
of the isospin asymmetry for a fixed nuclear density, $\rho=0.06~fm^{-3}$ and 
compares it to the QMC and Brueckner-Hartree-Fock (BHF) calculations. 
The relativistic models give 
larger fluctuation ratios than the corresponding value of $\rho_p/\rho_n$. 
We also observe that the fluctuation ratio in the present model is larger 
compared to the nonrelativistic BHF model. A pure mechanical disturbance 
would occur \cite{chomaz} 
if the instability preserves the ratio between protons and neutrons, i.e., 
$\frac{\delta\rho^-_p}{\delta\rho^-_n}=\frac{\rho_p}{\rho_n}$. Conversely if 
$\delta\rho^-_p=-\delta\rho^-_n$ then we should observe pure chemical 
disturbance. In the present case we observe that the disturbance along 
the unstable eigen direction conserves neither $\rho$ nor $y_p$ but has mixed 
character with both chemical and mechanical contents.

\subsection{Constrain on neutron star radii}
The symmetry energy plays an 
important role in describing the mass-radius relationship in neutron stars. 
Neutron stars are compact objects maintained by the equilibrium of 
gravity and the degenerecy pressure of the fermions together with a 
strong nuclear repulsion force due to the high density 
reached in their interior. The slope of the symmetry energy, $L$,  
constrains the neutron star radii. It is confirmed that the 
radii for the neutron stars with canonical mass $1.4 M_{\odot}$ 
are not affected by the 
symmetry energy at saturation density \cite{menezes}. However, 
in some cases the radii increase with $L$, while in others,
there is a decrease. In fact the radii are correlated
with a variation of the slope $L$. The radii increase  
up to a maximum value, then drop again. This behavior 
can be associated with a maximum theoretical value of $L$, 
and provide a possible constraint to nuclear matter. In the present model the
value of $L^0$ comes out to be $89$ MeV which is very close to the experimental 
observation \cite{tili}. 
The most direct connection between the astrophysical observations and the 
nuclear symmetry energy concerns neutron star radii ($R$) which are 
highly correlated with neutron star pressures near $\rho_0$. It is to be noted 
that Lattimer and Prakash \cite{constrain} found the radii of neutron stars 
for masses near the canonical mass $1.4 M_{\odot}$, obey a power-law relation:
\begin{equation}
R(M)=C(\rho,M)(P(\rho)/MeV fm^{-3})^{1/4}
\end{equation}
where $R(M)$ is the radius of a star, $P(\rho)$ is the pressure of neutron 
star matter at density $\rho$, and $C(\rho,M)$ is a constant for a given 
density and mass. Considering the structure of a neutron star as pure neutron 
matter, the value of this constant at quark mass $300$ MeV in 
our model comes out to be
\begin{equation}
C(2\rho_0,1.4M_{\odot})=5.90 km
\end{equation}
which is very near to $5.68 \pm 0.14$ km predicted by Lattimer {\it et al.} 
\cite{constrain}. It is to be noted that the mass of the neutron star for $pn$ 
matter with $\beta$ equilibrium 
comes out to be 2.7 $M_\odot$ with quark mass $300$ MeV and 1.64 $M_\odot$ with quark mass 
$40$ MeV. The details of such calculations incorporating the hyperons in the composition 
of neutron stars is in progress.
\section{Conclusion}
In the present work we have studied the EOS for asymmetric nuclear matter by 
using a modified quark-meson coupling model (MQMC). Self-consistent 
calculations were made by using a relativistic quark model with chiral 
symmetry along with the spurious 
center-of-mass correction, pionic correction for restoration of chiral symmetry, 
and short-distance correction for one-gluon exchange to realize different bulk 
nuclear properties. The instability in the two-component nuclear system is 
then analyzed. In asymmetric matter the isoscalar and isovector modes are 
coupled and the two separate inequalities for density oscillations and 
concentration oscillations no longer maintain a physical meaning for 
the selection of the nature of the instabilities. 

The symmetry energy, its slope $(L)$, and curvature parameter $(K_{sym})$ 
were found in reasonable agreement with experimental values. 
Without considering self 
interactions in the scalar field, we found
an analytic expression for the symmetry energy ${\cal E}_{sym}$
as a function of its slope $L$. Our result establishes a linear correlation
between $L$ and ${\cal E}_{sym}$. We also study the variation of correlation 
function $f(M_N^*,\rho_0,B_0,K_0)$ with the variation of quark masses.
The symmetry energy is 
correlated with neutron star radii. In this model we observe that, at 
twice the saturation density $(~0.3 fm^{-3})$, the constant $C(2\rho_0, M)$ 
is found 5.90 km in the canonical-mass region of $1.4 M_{\odot}$.
\section*{ACKNOWLEDGMENTS}
The authors would like to acknowledge the financial assistance from 
BRNS, India for the Project No. 2013/37P/66/BRNS.

\end{document}